\documentclass[12pt]{article}
\usepackage{amsmath,amssymb,graphicx,mathrsfs,slashed}
\usepackage[usenames, dvipsnames]{color} 
\usepackage{tabularx}
\newcommand{\be}{\begin{equation}}
        \newcommand{\ee}{\end{equation}}
\newcommand{\bea}{\begin{eqnarray}}
        \newcommand{\eea}{\end{eqnarray}}

\def\({\left(} \def\){\right)}

\begin{document}
\title{\vspace{-1.8in}
{Causality violations in Lovelock theories}}

\author{\large Ram Brustein, Yotam Sherf
\\
\vspace{-.5in}  \vbox{
\begin{center}
$^{\textrm{\normalsize
\ Department of Physics, Ben-Gurion University,
Beer-Sheva 84105, Israel}}$
\\ \small 
ramyb@bgu.ac.il,\ sherfyo@post.bgu.ac.il
\end{center}
}}
\date{}
\maketitle
\begin{abstract}
Higher-derivative gravity theories, such as Lovelock theories, generalize Einstein's general relativity (GR). Modifications to GR are expected when curvatures are near Planckian and appear in string theory or supergravity. But can such theories describe gravity on length scales much larger than the Planck cutoff length scale? Here we find  causality constraints on Lovelock theories that arise from the requirement that the equations of motion (EOM) of perturbations be hyperbolic. We find a general expression for the ``effective metric" in field space when Lovelock theories are perturbed around some symmetric background solution. In particular, we calculate explicitly the effective metric for a general Lovelock theory perturbed around cosmological Friedman-Robertson-Walker backgrounds and for some specific cases when perturbed around Schwarzschild-like solutions. For the EOM to be hyperbolic, the effective metric needs to be Lorentzian. We find that, unlike for GR, the effective metric is generically not Lorentzian when the Lovelock modifications are significant. So, we conclude that Lovelock theories can only be considered as perturbative extensions of GR and not as truly modified theories of gravity.   We compare our results to those in the literature and find that they agree with and reproduce the results of previous studies.
\end{abstract}
\newpage
\renewcommand{\baselinestretch}{1.5}\normalsize

\begin{subequations}
\renewcommand{\theequation}{\theparentequation.\arabic{equation}}
\section{Introduction}

General relativity (GR) is a  successful theory of gravity at large distance scales. At short distances one expects higher-derivative corrections to GR  in the form of derivatives and higher powers of curvature tensors as well as other modifications \cite{nojiri}. Unique among the generalizations of GR, the equations of motion (EOM) of Lovelock gravity  include at most second time derivatives \cite{ Lovelock:1971yv} (see \cite{Padmanabhan:2013xyr} for a review.) as do the Einstein equations.  General relativity in four spacetime dimensions can be viewed, from this perspective, as a special case of Lovelock gravity. The significance of having EOM which include at most second time derivatives and its relation to unitarity is explained in \cite{Brustein:2012he}.

Lovelock theories are relevant in several contexts. The quadratic Gauss-Bonnet (GB) theory  appears in the low energy limit of string theory \cite{Gross:1986iv,Zwiebach:1985uq} and in the study of higher dimensional black hole (BH) solutions \cite{Boulware:1985wk,Myers:1988ze,Cai:2003gr,Wheeler:1985nh} and cosmological solutions \cite{Wheeler:1985nh,Deruelle:1989fj}. More general terms were recently considered in \cite{Ciupke:2015msa,Farakos:2017mwd}. Studies of BH properties in the framework of the ADS/CFT correspondence also involve Lovelock gravity \cite{Cai:2003gr,Camanho:2009vw,Garraffo:2008hu}.

Here we would like to examine the possibility that Lovelock theories can introduce significant modification to GR also on large distance scales, so they can be viewed as  truly modified theories of gravity rather than provide just a small insignificant correction to GR \cite{Kiefer:2005uk,Camanho:2014apa}. Therefore, we will consider the case in which the coefficients of the higher derivative terms can be large, making the magnitude of the correction terms comparable to or larger than that of the Einstein term. We investigate the occurrence of causality violations in Lovelock theories by studying the hyperbolicity of the EOM of perturbations, as we explain in detail below.

A more general, related, method for determining causality violations  in Lovelock theories is the method of characteristics that uses the existence of a well-posed initial-value data to determine whether  perturbations propagate in a causal way. This method was used  in \cite{Izumi:2014loa,Reall:2014pwa} and more recently in \cite{Papallo:2015rna,Benakli:2015qlh} and in \cite{Papallo:2017qvl}. Previously, some related results were reviewed in \cite{Deruelle:2003ck}. In \cite{Papallo:2015rna} this method was used by Papallo and Reall to show that the effective metric for Schwarzschild-like solutions can change its signature near the horizon in the background of small black holes. Previously, similar results were also obtained in
\cite{LBH1,LBH2,LBH3,LBH4,LBH5,LBH6} by using a different method.  Our results are consistent with the existing results in the literature and our method allows us to find explicit numerical factors with ease. In \cite{Papallo:2017qvl} Papallo and Reall discussed the hyperbolicity of perturbation equations for Lovelock theories in the background of cosmological solutions and found the conditions for violations of hyperbolicity. We reproduce their results using the simpler method of calculating the effective metric.

Before addressing causality in Lovelock theories, we recall some simple  higher-derivative scalar field models. There, hyperbolicity of the EOM is equivalent to having a Lorentzian effective metric in field space. Then, after reviewing the results in the case of the scalar field models, we formulate along similar lines a general method to study causality in Lovelock theories by calculating the effective metric in field space. This method is  implemented for Lovelock theories expanded around cosmological Friedman-Robertson-Walker (FRW) backgrounds and around spherically symmetric BH solutions. The constraints indicate that the EOM of perturbations are not always hyperbolic when the Lovelock terms are significant.

\subsection{Review of causality constraints for scalar fields}\label{subsection1.1}

Causality constraints were extensively discussed, starting with \cite{Aharonov:1969vu} and later in \cite{ ArmendarizPicon:2005nz, Adams:2006sv, Bruneton:2006gf,Ellis:2007ic, Babichev:2007dw}.
To begin, we start by considering the addition of higher powers of derivatives to the lowest order scalar field Lagrangian in flat spacetime,
\begin{equation}
\mathcal{L}=\dfrac{1}{2}g^{\mu\nu}\nabla_\nu\phi \nabla_\mu\phi-\frac{1}{2}m^2\phi^2+ \sum_{n=2}\dfrac{1}{2n}\lambda_n\left(g^{\mu\nu}\nabla_\nu\phi \nabla_\mu\phi\right)^{n}
\label{Lphi1}~,
\end{equation}
where $\lambda_n$ are dimensionful coupling constants and $\nabla_{\mu}$ are covariant derivatives with respect to the background metric $g_{\mu\nu}$ and its associated Levi-Civita connection.
This form is chosen so the EOM contain at most second time derivatives. Defining $P(\phi)=g^{\mu\nu}\nabla_\nu\phi \nabla_\mu\phi $, and $F(P)=\sum\limits_{n=1}\frac{1}{n}\lambda_n P^{n}$, the Lagrangian takes the form $\mathcal{L}=\frac{1}{2}(F(P)-m^2\phi^2)$.

The EOM are given by
\begin{equation}
\left(2F''\nabla^\alpha\phi\nabla^\beta\phi+F'g^{\alpha\beta}\right) \nabla_\alpha\nabla_\beta\phi-m^2\phi=0
\label{Eoms1}~,
\end{equation}
which can be expressed as
\be
{G}^{\alpha\beta}\nabla_{\alpha}\nabla_{\beta}\phi -m^2 \phi=0\;.
\label{1.2}
\ee
The effective metric in field space is given by
\be
{G}^{\alpha\beta}= 2F''\nabla^\alpha\phi\nabla^\beta\phi+ F'g^{\alpha\beta}.
\label{1.3}
\ee
When the higher-derivative terms are absent, ${G}^{\alpha\beta}=g^{\alpha\beta}$. Since $g^{\alpha\beta}$ is a Lorentzian metric, the EOM are hyperbolic. However, when the higher derivative terms are included and when $\langle P(\phi) \rangle$ is non-vanishing, hyperbolicity is no longer automatic. We will encounter a similar phenomenon when discussing hyperbolicity of the EOM of Lovelock theories.

The hyperbolicity of the EOM is controlled by the effective metric ${G}^{\alpha\beta}$.  In general, a necessary condition for the hyperbolicity of the EOM is the Lorentzian structure of  ${G}^{\alpha\beta}$, which means that the effective metric has one negative eigenvalue for the time component and $D-1$ positive eigenvalues for the spatial components, or vice versa.

Obviously, when all eigenvalues are of the same sign, two successive events may be space-like separated and the traditional GR concepts of absolute future and past are not well defined. The evolution of solutions is said to be non-causal, which mathematically states that the Cauchy problem is not well posed.

Similarly, when the signs of the eigenvalues deviate from the standard pattern,  the local causal structure is modified. In extreme cases the contribution of the higher derivative terms may lead to the appearance of closed time-like curves. Since the lowest order effective metric is Lorentzian, the issue is whether the higher-order terms modify the effective metric in a substantial way.
In particular, in Eq.~(\ref{1.3}), the terms which are not necessarily proportional to the Lorentzian spacetime metric are proportional to $F''=\sum\limits_{n=2}(n-1)\lambda_n P^{n-2}$.

The couplings $\lambda_n$ in the Lagrangian~(\ref{Lphi1}) are dimensionful. So, for the higher order terms to make a significant contribution, the expectation value of  $\langle P(\phi) \rangle$ has to be large, such that the product $\lambda_n \langle P^n \rangle$ is large enough. In general, we expect terms of the form $P^n$ and also terms with more derivatives, such as $(\nabla_\alpha\nabla_\beta\phi)^n$ etc. to be significant at length scales near the cutoff length scale of the theory. At such scales, the hyperbolicity of the EOM is not a relevant concept because the semiclassical approximation breaks down. So to be viewed as a truly  higher-derivative theory, the corrections $\lambda_n P^n$ have to be large at length scales much larger than the cutoff length scale of the theory. Similar considerations will also apply to Lovelock theories.

\end{subequations}

\begin{subequations}
		\renewcommand{\theequation}{\theparentequation.\arabic{equation}}
\section{Causality violation in Lovelock theories}
\label{s2.0}

We will discuss Lovelock theories along the same lines of the discussion of the scalar field models. The analog of $\nabla_\alpha\phi\nabla_\beta\phi$ for the case of Lovelock theories is the Riemann tensor. The main complication in comparison to the scalar field models is the index structure and the gauge-redundancy in the Lovelock theories. We overcome these complications by a) using a formalism in which the EOM are written explicitly in terms of the Riemann tensor \cite{Iyer:1994ys,Brustein:2011gu} and b) studying gauge-invariant tensor perturbations around spherically symmetric spaces.

\subsection{Lovelock gravity}\label{s2.1}

Lovelock Lagrangians are defined as follows,
\begin{equation}
\mathcal{L}=\sum^{k_{max}}_{k=0}\lambda_k\mathcal{L}_k~,
\label{Ltot}
\end{equation}
where the sum runs up to $k_{max}\leq\frac{D-1}{2}$,   $\mathcal{L}_k$ are the $D$ dimensional Euler densities of order $k$, \footnote{In this paper we set $\lambda_0=0 $ and choose units in which $ \lambda_{1}=1$.}
\begin{equation}
\mathcal{L}_k~=~\dfrac{1}{2^k}\delta^{aba_1b_1...a_kb_k}_{cdc_1d_1...c_kd_k} \mathcal{R}^{~~cd}_{ab}\mathcal{R}^{~~~c_1d_1}_{a_1b_1}\cdots \mathcal{R}^{~~~c_kd_k}_{a_kb_k}~.
\label{Lk}
\end{equation}
The tensors $\mathcal{R}^{~~c_1d_1}_{a_1b_1}$ are the background Riemann tensors and the tensor $\delta$ in Eq.~(\ref{Lk}) is  fully antisymmetric in its upper and lower indices and has the same symmetry properties of the Riemann tensor in each set of four indices. This tensor can be expressed as the following product of Kronecker delta's
\begin{equation}
\delta^{a_1b_1a_2b_2}_{c_1d_1c_2d_2} \cdots^{a_kb_k}_{c_kd_k}= ~\delta^{ {\large\pmb[} a_1}_{c_1}\delta^{b_1}_{d_1}\cdots \delta^{a_k}_{c_k}\delta^{a_k {\large\pmb]}}_{c_k}= ~\delta^{a_1}_{{\large\pmb[} c_1}\delta^{b_1}_{d_1}\cdots \delta^{a_k}_{c_k}\delta^{a_k}_{c_k{\large\pmb]}}~.
\label{delta1}
\end{equation}
The variation of a single $\mathcal{L}_k$  generalizes the Einstein tensor,
\begin{equation}
(\mathcal{G}^p_{~q})_k~=~ -\dfrac{1}{2^{k+1}}\delta^{paba_1b_1...a_kb_k}_{qcdc_1d_1...c_kd_k} \mathcal{R}^{~~cd}_{ab} \mathcal{R}^{~~~c_1d_1}_{a_1b_1}\cdots\mathcal{R}^{~~~c_kd_k}_{a_kb_k}~.
\label{EOMk}
\end{equation}
One can check that the $k=1$ term gives the standard Einstein tensor,
\begin{equation}
	\begin{split}
		(\mathcal{G}^p_{~q})_1~&=~ -\dfrac{1}{4}\delta^{pab}_{qcd} \mathcal{R}^{~~cd}_{ab} ~\\ &=~-\dfrac{1}{4}\left(\delta^p_q\delta^{ab}_{cd}-\delta^p_c\delta^{ab}_{qd}+ \delta^p_d\delta^{ab}_{qc}\right) \mathcal{R}^{~~cd}_{ab}\\& =~~\mathcal{R}^p_q-\dfrac{1}{2}\delta^p_q\mathcal{R}~.
		\label{EOMk1}
	\end{split}
\end{equation}
The full generalized Einstein tensor is given by
\be
\mathcal{G}^p_{~q} = \sum_k\lambda_k (\mathcal{G}^p_{~q})_k.
\label{ETGeneral}
\ee

The coefficients $\lambda_k$ are generically dimensionful. They can be expressed, in units in which $\lambda_1=1$, as $\lambda_k= \widetilde{\lambda}_k (\ell_k)^{2(k-1)}$ for $k\geq 2$ and  $\widetilde{\lambda}_k$ are dimensionless numerical coefficients  of order unity. The parameters $\ell_k$ have the dimension of length and they determine at which length scales the corrections are important. Without any tuning the dominant contribution of the correction terms becomes significant when approaching the ultraviolet length scale. To act as truly modified theories of gravity, the coupling of the higher-order terms have to be anomalously large, or alternatively, the length scales $\ell_k$ have to be anomalously large, so that the higher-order terms can affect the solutions of the EOM at length scales much larger than the cutoff length scale around the Lorentz-invariant vacuum.

We would like to show that when the higher-order couplings are large, the EOM are no longer guaranteed to be hyperbolic. So, the theory cannot be viewed as a consistent theory.

\subsection{Effective metric for propagation of perturbations}
\label{s2.2}

The purpose of this section is to find the effective metric in field space for Lovelock theories. We later require that the metric be Lorentzian, so the EOM of perturbations is hyperbolic. This metric is also sometimes called the ``acoustic metric" \cite{Babichev:2007dw}.

We start by expanding the metric ${g}_{ab}$ about the  background $\bar{g}_{ab}$,
\be
{g}_{ab}=\bar g_{ab}+ h_{ab}
\label{backgrounddexpansion}
\ee
and then we expand the generalized Einstein tensor in Eq.~(\ref{ETGeneral}) to first order in $h_{ab}$. We  identify the effective metric by finding all the kinetic terms -- terms of the form $\nabla\nabla h$.

A straightforward naive variation of Eq.~(\ref{ETGeneral}) is not sufficient, since, as we shall see below, mass terms may be disguised as  kinetic terms. Therefore we must identify and isolate these mass terms in the expansion of  Eq.~(\ref{ETGeneral}). Fortunately, in Lovelock theories the identification of the kinetic terms is made easier  by expressing $(\mathcal{G}^{pq})_k$ as follows  \cite{Iyer:1994ys},
\begin{equation}
\mathcal{G}^{pq}~=~\mathcal{X}^{pabc}\mathcal{R}^q_{~abc}-\dfrac{1}{2}g^{pq}\mathcal{L}.
\label{EOM1}
\end{equation}
The tensor $\mathcal{X}^{pabc}\equiv\frac{\partial\mathcal{L}}{\partial\mathcal{R}_{pabc}}$ has the symmetry properties of the Riemann tensor and is given by a sum $\mathcal{X}=\sum\limits_k \lambda_k \mathcal{X}_k$, with
\begin{equation}
(\mathcal{X}_k)^{pq}_{~rs}~=~\dfrac{k}{2^k}~\delta^{pqa_2b_2}_{rsc_2d_2} \cdots^{a_kb_k}_{c_kd_k}\mathcal{R}^{~~~c_2d_2}_{a_2b_2}\cdots\mathcal{R}^{~~~c_kd_k}_{a_kb_k}~.
\label{Chi1}
\end{equation}
The tensor $\mathcal{G}^{pq}$ contains a  term proportional to the background metric $ g^{pq} $  which does not contribute to the effective metric. The variation of this term, for some graviton polarizations, could sometime result in a mass term which is disguised as a kinetic term. But, as we show below, these mass terms can be systematically removed  when specific graviton polarizations are considered.

The tensor $\mathcal{X}\mathcal{R}$  contains second derivatives of the metric and we will show below that its variation will allow us to identify the effective metric. Technically, due to symmetry considerations, it is easier to extract the effective metric from the combination $\mathcal{G}^p_{~q}+\frac{1}{2}\delta^p_{q}\mathcal{L}$  rather than from  $\mathcal{X}^{pabc}\mathcal{R}_{qabc}$ directly.

We begin by defining some useful relations.
The first order expansion of the Riemann tensor is given by,
\begin{equation}
\delta\mathcal{R}^{(1)}_{abcd}~=~\dfrac{1}{2}\big(\nabla_a\nabla_c h_{bd}+\nabla_b\nabla_{d} h_{ac}-\nabla_a\nabla_d h_{bc}-\nabla_b\nabla_c h_{ad}\big)~.
\label{Riem1}
\end{equation}
The commutation relations of covariant derivatives are expressed in terms of the Riemann tensor,
\begin{equation}
\left[\nabla_c, \nabla^p\right]h^c_q=-\mathcal{R}_{c~~\alpha}^{~pc}h^{\alpha}_q-\mathcal{R}^{~p~\alpha}_{c~q}h^c_{\alpha}.
\label{Rcr}
\end{equation}	
Here $h^c_q$ is a symmetric tensor which later will be viewed as the tensor perturbation around some background solution.

The contraction of the delta tensor with $\delta\mathcal{R}^{(1)}$ yields the useful relation
\begin{equation}
\delta^{aba_1b_1\dots a_kb_k}_{cdc_1d_1\dots c_k d_k}\delta\mathcal{R}^{~~cd}_{ab}~=~2\delta^{aba_1b_1\dots a_kb_k}_{cdc_1d_1\dots c_k d_k}\nabla_b\nabla^{d} h_{a}^{c}.
\label{Riemc}
\end{equation}

The contribution of the order-$k$ Lovelock term to the kinetic part -- the part which contains terms of the form $\nabla\nabla h$ -- is obtained by varying $\left(\mathcal{G}^p_{~q}+\frac{1}{2} \delta^p_{q}\mathcal{L}\right)_k$,
\begin{align}
&\delta(\mathcal{G}^p_{~q}+\frac{1}{2} \delta^p_{q}\mathcal{L})_k~=~	 \\ -\dfrac{k}{2^{k}} & \!\left(\delta^{paba_1b_1...a_kb_k}_{qcdc_1d_1...c_kd_k} -\delta^p_q\delta^{aba_1b_1...a_kb_k}_{cdc_1d_1...c_kd_k}\right)\mathcal{R}^{~~~c_1d_1}_{a_1b_1} \cdots\mathcal{R}^{~~~c_{k}d_{k}}_{a_{k}b_{k}}\nabla_b\nabla^{d} h_{a}^{c}~. \nonumber
\label{EomV}
\end{align}
Now, let us  define the following tensor,
\begin{equation}
\left(\mathcal{T}^{pab}_{~~qcd}\right)_k~ =~\dfrac{k}{2^k}~\delta^{paba_1b_1...a_kb_k}_{qcdc_1d_1...c_kd_k}\mathcal{R}^{~~c_1d_1}_{a_1b_1} \cdots\mathcal{R}^{~~c_{k}d_{k}}_{a_{k}b_{k}}\;,
\label{Tkin}
\end{equation}
then, using Eq.~(\ref{Chi1}), the kinetic part  becomes
\begin{equation}
\delta\left(\mathcal{G}^p_{~q}+\frac{1}{2} \delta^p_{q}\mathcal{L}\right)= - \sum_{k=1}^{k_{max}}\lambda_k (\mathcal{T}^{pab}_{~~qcd}-\delta^p_q\mathcal{X}^{ab}_{~cd})_k\nabla_b\nabla^{d} h_{a}^{c}~.
\label{kinctot}
\end{equation}

We evaluate the effective metric here for tensor perturbations around spherically symmetric solutions of the EOM. We do so for simplicity and to facilitate the comparison to GR \cite{Gibbons:2002pq}. Similar equations can also be obtained for scalar and vector perturbations. Gauge-invariant tensor perturbations around spherically symmetric backgrounds are  transverse and traceless,
\begin{eqnarray}
h^i_{~i}, ~\nabla_ih^i_{~j}=0.
\label{Gauge1}
\end{eqnarray}
The ability to choose transverse-traceless (TT) perturbations about maximally symmetric subspaces relies on the geometric properties of these spaces and not on the gravitational action.  This is discussed, for example, in \cite{Higuchi:1986wu}, where it is shown that tensor perturbations about a maximally symmetric subspace can always be defined  as TT.

We first show how this process works in the simplest case of GR, corresponding to the $k=1$ term in Eq.~(\ref{kinctot}). Then, we repeat the process for the GB theory. We extract the relevant kinetic part by using the commutation relations in Eq.~(\ref{Rcr}). Finally, we write the explicit expression for the effective metric for tensor perturbations for the general Lovelock theory.

We start by expressing the delta tensor in terms of lower-order delta tensors,
\begin{equation}
\begin{split}
\left(\mathcal{T}^{pab}_{~~qcd}-\delta^p_q\mathcal{X}^{ab}_{~cd}\right)_1\nabla_b\nabla^{d} h_{a}^{c}~&=~\dfrac{1}{2}	
	\big(\delta^{pab}_{qcd}-\delta^p_q\delta^{ab}_{cd}\big)\nabla_b\nabla^{d} h_{a}^{c}\\=~\dfrac{1}{2} &
\big(	\delta^a_q\delta^{pb}_{cd}-	\delta^a_c\delta^{pb}_{qd}+	\delta^a_d\delta^{pb}_{qc}-\delta^p_q\delta^{ab}_{cd}\big)\nabla_b\nabla^{d} h_{a}^{c}.
\label{Einsvar1}
\end{split}
\end{equation}
For tensor perturbations, using the commutation relations,
\begin{equation}
\begin{split}
\left(\mathcal{T}^{pab}_{~~qcd}-\delta^p_q\mathcal{X}^{ab}_{~cd}\right)_1\nabla_b\nabla^{d} h_{a}^{c} = &\dfrac{1}{2}\left(\delta^a_q\delta^p_cg^{bd}\nabla_b\nabla^{d} h_{a}^{c}-\nabla_c\nabla^{p} h_{q}^{c}\right)~\\&=~\dfrac{1}{2}\left(\delta^a_q\delta^p_cg^{bd}\nabla_b\nabla^{d} h_{a}^{c}+\mathcal{R}_{c~~\alpha}^{~pc}h^{\alpha}_q+\mathcal{R}_{c~q}^{~p~\alpha}h^{c}_{\alpha}\right)~.
	\end{split}
\end{equation}

In this form, kinetic terms and mass terms can be separated.  Now, dismissing  mass terms while keeping the kinetic terms, we obtain
\begin{equation}
\left(\mathcal{T}^{pab}_{~~qcd}-\delta^p_q\mathcal{X}^{ab}_{~cd}\right)_1= \dfrac{1}{2}\delta^a_q\delta^p_cg^{bd}\nabla_b\nabla_{d} h_{a}^{c} + \textrm{mass\ terms}.
\label{Einsvar2}
\end{equation}

Next, we need to collect all contributions for a  specific graviton polarization $h_{a}^{c} $.
We choose the polarization by fixing $a$ and $c$, which implies that these indices are not summed over. To avoid confusion we label them as ${\widetilde{\textbf{a}}}$ and ${\widetilde{\textbf{c}}}$. In general, one has to diagonalize the effective metric. However, if the background solution has enough symmetry, the effective metric is already diagonal, as is the case in Eq.~(\ref{Einsvar2}).

Now, the effective metric can be identified in a covariant form
\be
\left[G^{b{d}}\right]^{\widetilde{\textbf{a}}}_{\widetilde{\textbf{c}}} =\dfrac{1}{2} \delta^{{\widetilde{\textbf{a}}}}_{q}\delta^p_{{\widetilde{\textbf{c}}}}g^{bd}
=\dfrac{1}{2}g^{bd}.
\ee

So, for GR, the effective metric for all polarizations is equal to the spacetime metric $G^{bd}=g^{bd}$. Hence, $G^{bd}$ is  a Lorentzian metric. This results in hyperbolic EOM for any choice of graviton polarization.

When considering general Lovelock theories,  the effective metric can be different for different graviton polarizations.  This happens because higher order terms ($k\geq2$) include explicitly the Riemann tensor. For example, a term like $\delta^{\widetilde{\textbf{a}}}_q\mathcal{R}^{pb}_{~\widetilde{\textbf{c}}d}$ can have a polarization-dependent contribution if different components of the Riemann tensor take on different background values.

Next, we demonstrate our method for the Gauss-Bonnet term, $k=2$. We evaluate the effective metric (using X'act and X'pert Mathematica package \cite{Brizuela:2008ra}), repeating the process described previously, where the contribution of the second order Lovelock term together with the Einstein term is Eq~(\ref{kinctot}),
\begin{equation}
\begin{split}
&\sum_{k=1,2}	\lambda_k\dfrac{k}{2^k}(\mathcal{T}^{pab}_{~~qcd}- \delta^p_q\mathcal{X}^{ab}_{~cd})_k\nabla_b\nabla^{d} h_{a}^{c} =
\dfrac{1}{2}\left[\big(\delta^{pab}_{qcd}-\delta^p_q\delta^{ab}_{cd}\big)\right. \\& \left.+~ \lambda_2\big(\delta^{paba_1b_1}_{qcdc_1d_1}\mathcal{R}^{~~~c_1d_1}_{a_1b_1} -\delta^p_q\delta^{aba_1b_1}_{cdc_1d_1} \mathcal{R}^{~~~c_1d_1}_{a_1b_1}\big)\right]\nabla_b\nabla^{d} h_{a}^{c}.
\label{GBsvar}
\end{split}
\end{equation}
Choosing tensor perturbations, using commutation relations of covariant derivatives to dispose of mass terms and fixing the indices  $a$, $c$, we find the effective  metric
$\left[G^{{b}}_{{\ d}}\right]^{\widetilde{\textbf{a}}}_{\widetilde{\textbf{c}}}$,
\begin{align}
\label{GBvar1}
&\left[G^{b}_{{\ d}}\right]^{\widetilde{\textbf{a}}}_{\widetilde{\textbf{c}}} \nabla_b\nabla^{d} h_{{\widetilde{\textbf{a}}}}^{{\widetilde{\textbf{c}}}}~=~
 \cr &
\dfrac{1}{2} \Bigg[(\delta^{\widetilde{\textbf{a}}}_q\delta^p_{\widetilde{\textbf{c}}} \delta^{b}_{d})+	2\lambda_2\delta^{\widetilde{\textbf{a}}}_q\big(\delta^b_d (\mathcal{R}\delta^p_{\widetilde{\textbf{c}}} -2\mathcal{R}^p_{\widetilde{\textbf{c}}})+2\delta^p_d\mathcal{R}^b_{\widetilde{\textbf{c}}} -\delta^p_{\widetilde{\textbf{c}}}\mathcal{R}^b_d +\mathcal{R}^{pb}_{~{\widetilde{\textbf{c}}}d}\big)
\cr &
+2\big(\delta^b_d(\mathcal{R}^{p{\widetilde{\textbf{a}}}}_{~q{\widetilde{\textbf{c}}}} -\delta^p_q\mathcal{R}^{\widetilde{\textbf{a}}}_{\widetilde{\textbf{c}}})\big)  +\delta^b_q\mathcal{R}^{p{\widetilde{\textbf{a}}}}_{~{\widetilde{\textbf{c}}}d}~ +\delta^p_d\big(\delta^b_q\mathcal{R}^{\widetilde{\textbf{a}}}_{\widetilde{\textbf{c}}} +\mathcal{R}^{{\widetilde{\textbf{a}}}b}_{~q{\widetilde{\textbf{c}}}}\big)
\\ &
+\delta^p_{\widetilde{\textbf{c}}}\big(\delta^b_d\mathcal{R}^{\widetilde{\textbf{a}}}_q -\delta^b_q\mathcal{R}^{\widetilde{\textbf{a}}}_d -\mathcal{R}^{{\widetilde{\textbf{a}}}b}_{~qb}\big)  +\delta^p_q\mathcal{R}^{ab}_{~{\widetilde{\textbf{c}}}d} -\delta^p_q(\mathcal{R}^{{\widetilde{\textbf{a}}}b}_{~{\widetilde{\textbf{c}}}d} -\delta^b_d\mathcal{R}^{{\widetilde{\textbf{a}}}}_{{\widetilde{\textbf{c}}}})\Bigg]
\nabla_b\nabla^{d} h_{{\widetilde{\textbf{a}}}}^{{\widetilde{\textbf{c}}}}. \nonumber
\end{align}
Recall that we are considering background solutions in which the effective metric is diagonal. Then, $p= \widetilde{\textbf{c}}$ and $q=\widetilde{\textbf{a}}$.

We can now determine whether the metric $G^{bd}$ is Lorentzian.
If the Riemann tensor is non-vanishing on a solution, then, in general, $G^{bd}$ will not be proportional to the spacetime metric $g^{bd}$ and therefore the hyperbolicity of the EOM for a specific choice of polarization is not guaranteed.

In summary, the extraction of the effective metric is as follows. Start from Eq.~(\ref{kinctot}). Express the higher Lovelock terms in terms of  delta tensors. The result is the following,
\begin{align}
& -\sum_{k=1}^{k_{max}}\lambda_k (\mathcal{T}^{pab}_{~~qcd}-\delta^p_q\mathcal{X}^{ab}_{~cd})_k\nabla_b\nabla^{d} h_{a}^{c}
\cr
& = -\sum_{k=1}^{k_{max}}\lambda_k \dfrac{k}{2^k}\Biggl( \delta^{paba_1b_1...a_kb_k}_{qcdc_1d_1...c_kd_k}\mathcal{R}^{~~~c_1d_1}_{a_1b_1} \cdots\mathcal{R}^{~~~c_{k}d_{k}}_{a_{k}b_{k}}
\cr & \hspace{1in}
-\delta^p_q~\delta^{aba_2b_2}_{cdc_2d_2} \cdots^{a_kb_k}_{c_kd_k}\mathcal{R}^{~~~c_2d_2}_{a_2b_2} \cdots\mathcal{R}^{~~~c_kd_k}_{a_kb_k}\Biggr) \nabla_b\nabla^{d} h_{a}^{c}~.
\label{kingen1}
\end{align}
Then, we choose tensor perturbations and use the commutation relations of covariant derivatives in order to isolate the kinetic terms, dropping the mass terms. Third, choose the polarization  ${\widetilde{\textbf{a}}}$ and ${\widetilde{\textbf{c}}}$. Since the Einstein result ($k=1$) is known to be $\delta^{b}_{~d}$,
\begin{align}
\big[G^{{b}}_{~{d}}\big]^{\widetilde{\textbf{a}}}_{\widetilde{\textbf{c}}} \nabla_b\nabla^{d} h_{{\widetilde{\textbf{a}}}}^{{\widetilde{\textbf{c}}}}
& =\Biggl[ \delta^{b}_{~d}-\sum_{k=2}^{k_{max}}\lambda_k \dfrac{k}{2^k}\Biggl( \delta^{p\widetilde{\textbf{a}}ba_1b_1...a_kb_k}_{q\widetilde{\textbf{c}}dc_1d_1...c_kd_k} \mathcal{R}^{~~c_1d_1}_{a_1b_1} \cdots\mathcal{R}^{~~~c_{k}d_{k}}_{a_{k}b_{k}}
\cr &
-\delta^p_q~\delta^{\widetilde{\textbf{a}}ba_2b_2}_{\widetilde{\textbf{c}}dc_2d_2} \cdots^{a_kb_k}_{c_kd_k}\mathcal{R}^{~~~c_2d_2}_{a_2b_2}\cdots\mathcal{R}^{~~~c_kd_k}_{a_kb_k}\Biggr) \Biggr] \nabla_b\nabla^{d} h_{{\widetilde{\textbf{a}}}}^{{\widetilde{\textbf{c}}}}~.
\label{kingen2}
\end{align}
Finally, set $p= \widetilde{\textbf{c}}$ and $q=\widetilde{\textbf{a}}$.
Now, the effective metric can be read off as the symmetric tensor contracting the second derivative terms acting on $h_{{\widetilde{\textbf{a}}}}^{{\widetilde{\textbf{c}}}}$. The complexity of the expression increases rapidly with $k$, which means that, in practice, getting the explicit expressions is quite complicated.

The general expression for the effective metric will be a polynomial of a higher degree in the various curvature tensors, or alternatively, a multinomial in the metric components and their first and second derivatives (Higher than second derivatives do not appear.).  When the background curvature vanishes, or is perturbatively small (in a sense that will be clarified shortly), then the higher order terms add a small correction to the leading Lorentzian Einstein term. However, when the higher order terms are as important as the Einstein term, the Lorentzian nature of the metric is no longer guaranteed. In general, only under very special circumstances, the metric is indeed Lorentzian. We show this explicitly by studying the expansion of the higher order Lovelock theories around some non-trivial background solutions in the next section.

\end{subequations}

\begin{subequations}
	\renewcommand{\theequation}{\theparentequation.\arabic{equation}}
\section{Explicit calculations of the effective metric }
 \subsection{Effective metric in a cosmological Friedman-Robertson-Walker spacetime}
 \label{s3.1}

Consider the case of a D-dimensional homogeneous and isotropic Friedman-Robertson-Walker (FRW) spacetime  whose line element is given by
\begin{equation}
ds^2~=~-dt^2+a(t)^2\gamma_{{i}{j}}dx^{{i}}dx^{{j}},
\label{frw}
\end{equation}
where $i,j=1,2,..,D-1$ denote spatial components. Here we choose for simplicity $\gamma_{{i}{j}}=\delta_{ij}$. This type of spacetime solves the Lovelock EOM in the presence of matter \cite{Wheeler:1985nh,Deruelle:1989fj}. We will not need the  detailed descriptions of the solutions for a general Lovelock theory which can be found in \cite{Deruelle:1989fj}, nor will we need the detailed description of the matter sources, which can also be found there.

We just need to know that the solutions exist and the sources are physical. Particularly relevant is the existence of solutions that describe a universe undergoing decelerated expansion, similar to matter-dominated or radiation dominated solutions of Einstein's equations. We restrict our attention to dimensions higher than 4 for simplicity.
It would be interesting to study the hyperbolicity of these cosmological solutions upon compactification to 4D. Dimensional reduction of Lovelock theories was discussed in detail in \cite{MuellerHoissen:1989yv,MuellerHoissen:1985mm} and more recently in \cite{VanAcoleyen:2011mj}. In  some cases such dimensional reduction may lead to a scalar-graviton coupling. A detailed discussion of the conditions of hyperbolicity in various compactification schemes is outside the scope of this paper.

The non-vanishing components of the Riemann tensor are the following,
 \begin{eqnarray}
 \mathcal{R}^{ij}_{kl}&=& H^2\delta^{ij}_{kl},
 \label{Rimfrw12}
 \\
 \mathcal{R}^{ti}_{tj}&=&\delta^{i}_{j}\frac{\ddot{a}}{a}=\delta^{i}_{j}(H^2+\dot H)
 \label{Rimfrw1},
 \end{eqnarray}
where $i,j,k,l=1,2,..,D-1$ denote, again, the spatial components,  $H^2=(\dot{a}/a)^2$ is the Hubble parameter and $\ddot{a}/a=H^2+\dot{H}$. Here, a dot represents a time derivative. The tensor perturbations are defined as follows,
\begin{subequations}
	\begin{align}
h^i_{~i} & =0~,
\label{g1}
\\
	h^i_{~t}& =0~,
\label{g2}
\\
\nabla_ih^i_{j}& =0.
	\label{g3}
	\end{align}
\end{subequations}

We now wish to explain how to evaluate the effective metric for perturbations around an FRW solution, for a general  Lovelock theory. To illustrate the procedure, we  first evaluate the metric for the Gauss-Bonnet term ($k=2$) and then calculate the general expression for the metric.

The contribution of the GB term to the metric is the following,
\begin{equation}
	\dfrac{1}{2}\lambda_2(\mathcal{T}^{pab}_{~~qcd}-\delta^p_q\mathcal{X}^{ab}_{~cd})_2~=~\dfrac{1}{2}	\lambda_2\big(\delta^{paba_1b_1}_{qcdc_1d_1}\mathcal{R}^{~~~c_1d_1}_{a_1b_1}-\delta^p_q\delta^{aba_1b_1}_{cdc_1d_1}\mathcal{R}^{~~~c_1d_1}_{a_1b_1}\big)~.
	\label{eff-fr}
\end{equation}
The computation of the effective metric is carried out by using the following  identities,
\begin{gather}
	\delta^{a_1b_1... a_kb_ka_{k+1}b_{k+1}... a_nb_n}_{c_1d_1... c_kd_ka_{k+1}b_{k+1}... a_nb_n}~=~\dfrac{(D-k)!}{(D-2n)!}~\delta^{a_1b_1... a_kb_k}_{c_1d_1... c_kd_k}~,\\
		\delta^{tpaa_1b_1\dots a_kb_k}_{tqcc_1d_1\dots c_kd_k}~=~\delta^{\bar{p}\bar{a}\bar{a_1}\bar{b_1}\dots \bar{a_k}\bar{b_k}}_{\bar{q}\bar{c}\bar{c_1}\bar{d_1}\dots \bar{c_k}\bar{d_k}}~,
			\label{DelId}
\end{gather}
where barred indices $~\bar{p}, \bar{q}=1,2,\dots,D-1$, etc.,  denote spatial components. The notation employed  in Eq.~(\ref{DelId}) indicates that the left-hand side vanishes unless all but the $t$-indices are spatial.

We begin by computing $G^t_{~t}$, setting $b,~d=t$ in Eq.~(\ref{eff-fr}),
\begin{equation}
\begin{split}
(\delta^{pata_1b_1}_{qctc_1d_1}-\delta^p_q\delta^{ata_1b_1}_{ctc_1d_1})\mathcal{R}^{~~~c_1d_1}_{a_1b_1} &=(\delta^{\bar{p}\bar{a}\bar{a_1}\bar{b_1}}_{\bar{q}\bar{c}\bar{c_1}\bar{d_1}} -\delta^p_q\delta^{\bar{a}\bar{a_1}\bar{b_1}}_{\bar{c}\bar{c_1}\bar{d_1}}) \mathcal{R}^{~~~\bar{c_1}\bar{d_1}}_{\bar{a_1}\bar{b_1}}
\\
=2H^2(D-3)&\left[\delta^{\bar{p}\bar{a}}_{\bar{q}\bar{c}}(D-4)-\delta^{p}_{q} \delta^{\bar{a}}_{\bar{c}}(D-2)\right].
\label{frTT}
\end{split}
\end{equation}
For the spatial components $G^{\bar{p}}_{~\bar{q}}$, we have contributions from the two non-vanishing Riemann components.  The contribution from the purely spatial components of the  Riemann Eq.~(\ref{Rimfrw12}) is given by
\begin{align}
&\left[\delta^{\bar{p}iji_1j_1}_{\bar{q}klk_1l_1} -\delta^p_q\delta^{iji_1j_1}_{klk_1l_1}\right]\mathcal{R}^{~~~k_1l_1}_{i_1j_1}  =
\cr &
2H^2(D-4)\left[\delta^{\bar{p}ij}_{\bar{q}kl}(D-5)-\delta^p_q\delta^{ij}_{kl}(D-3)\right]
\label{frS1}
\end{align}
and the contribution of the mixed time-space components in Eq.~(\ref{Rimfrw1}) reads,
\begin{gather}
4\left(\delta^{\bar{p}ijtj_1}_{\bar{q}kltl_1}- \delta^p_q\delta^{ijtj_1}_{kltl_1}\right)\mathcal{R}^{~~~tl_1}_{t\bar{j_1}}= 4(H^2+\dot{H})\left[\delta^{\bar{p}ij}_{\bar{q}kl}(D-4)-\delta^p_q\delta^{ij}_{kl}(D-3)\right]~.
\label{frS2}
\end{gather}
The total spatial contribution comes from combining Eqs.~(\ref{frS1}) and (\ref{frS2}).

We proceed in evaluating the spatial part of the effective metric. We add the Einstein term contribution from Eq.~(\ref{Einsvar1}) to that of  Eq.~(\ref{eff-fr}), and then impose the gauge conditions and use the commutation relations of covariant derivatives, subtract the mass terms and fix the polarization indices ${\widetilde{\textbf{i}}},~{\widetilde{\textbf{j}}}$. The result is
\begin{align}
{\big[G^{tt}\big]^{\widetilde{\textbf{i}}}_{\widetilde{\textbf{j}}} }& ~=~ g^{tt}~\left[1+2\lambda_2 H^2(D-3)(D-4)\right]~.
	\label{frTTg1}
\\
{\big[G^{\bar{b}\bar{d}}\big]^{\widetilde{\textbf{i}}}_{\widetilde{\textbf{j}}} }
&~=~\sum_{p,q=1}^{D}	 \delta^{{\widetilde{\textbf{i}}}}_{\bar{q}}\delta^{\bar{p}}_{{\widetilde{\textbf{j}}}} g^{\bar{b}\bar{d}}\left[1+\lambda_2\big(2H^2(D-4)(D-5)+4(H^2+\dot{H})(D-4)\big)\right]
\cr &~=
~g^{\bar{b}\bar{d}}\left[1+2 \lambda_2 H^2(D-3)(D-4)\left(1+\dfrac{2}{D-3}\dfrac{\dot H}{H^2}\right)\right]~.
\label{frSg1}
\end{align}
Symmetry leads to the same effective metric for all polarizations. We also note that in this case the effective metric is not proportional to the spacetime metric $g^{\mu\nu}$. Rather the time-time component and the space-space component are proportional to the respective spacetime metric components with different proportionality factors.

As previously mentioned, a necessary condition for  hyperbolic EOM is that the effective metric be Lorentzian. Thus, our interest is in the conditions under which the metric ${\left[G^{\mu\nu}\right]^{\widetilde{\textbf{i}}}_{\widetilde{\textbf{j}}} }~$ is either Lorentzian or non-Lorentzian. To be Lorentzian, the factor multiplying $g^{tt}$ in Eq.~(\ref{frTTg1}) and the factor multiplying $g^{\bar{b}\bar{d}}$ in Eq.~(\ref{frSg1}) have to have the same sign. Conversely, if these factors have a different sign then the effective metric will not be Lorentzian.

Let us first assume that $\lambda_2$ is positive (The case in which $\lambda_2$ is negative is problematic as we explain below.). Then, the time-time component of the metric $G^{tt}$ is always negative. So, the determining factor is the  sign of $G^{\bar{b}\bar{d}}$. When $2 \lambda_2 H^2(D-3)(D-4)\left(1+\dfrac{2}{D-3}\dfrac{\dot H}{H^2}\right)<-1$, the metric is not Lorentzian. This happens under two conditions: (i) that $\lambda_2 H^2$ is large, $\lambda_2 H^2\gtrsim 1$ and (ii) that $\dfrac{\dot H}{H^2}$ is negative, such that $1+\dfrac{2}{D-3}\dfrac{\dot H}{H^2}<0$.

To understand the significance of the conditions, let us consider a solution of decelerated expansion of the form $a(t)\sim t^\alpha$ with $0<\alpha<1$. In this case, $\dot H = - H^2/\alpha $. So,  the effective metric is not Lorentzian when $1+2 \lambda_2 H^2(D-3)(D-4)\left(1-\dfrac{2}{(D-3)\alpha}\right)<0$.  This condition (when $\lambda_2 H^2\gtrsim 1$) rules out most of the parameter space of decelerated expanding isotropic solutions found in \cite{Camanho:2015yqa}.

For example, for a radiation dominated universe in $D$ dimensions, $\alpha=2/D$ and a matter domination universe corresponds to $\alpha=2/(D-1)$. The exact numerical conditions involve additional $D$-dependent factors which can be worked out for any desired specific case. For example, for $D=5$, the effective metric is not Lorentzian for the whole range $0<\alpha<1$ when the correction terms are significant $\lambda_2 H^2\gtrsim 1$. We will not list here all the cases, as it is by now clear that the effective metric is non-Lorentzian for many of them.

A way to interpret our results is the following. To ensure hyperbolicity one can simply demand that $\lambda_2 H^2<1$. This means that the cutoff scale of the theory is set by the correction term such that it is subdominant. Alternatively, one can allow $\lambda_2 H^2$ to be large, but then one has to impose conditions on $\dot{H}$, which again can be interpreted as imposing a cutoff on the theory such that some part of the corrections is subdominant.

The case $\lambda_2<0$ is problematic from several perspectives. When one considers black hole solutions, as we discuss later, one finds that negative $\lambda_2$ can lead to  naked singularities \cite{Zwiebach:1985uq}. From our perspective, a negative $\lambda_2$ means that if $H$ is small $G^{tt}$ is negative, while if $H$ is large enough $G^{tt}$ is positive. This suggests that for consistency and to allow solutions with small $H$, one needs to impose $|\lambda_2 H^2|<1/(2(D-3)(D-4) )$, which effectively sets the cutoff of the theory at this scale.
Setting these issues aside, we can analyze the case $\lambda_2<0$ along the same lines as we did in the case $\lambda_2>0$. Here if $|\lambda_2| H^2\gtrsim 1$ and $1+\dfrac{2}{D-3}\dfrac{\dot H}{H^2}<0$, we find that the effective metric is non-Lorentzian.

We now turn to discuss the results for the effective metric for  general Lovelock theories.

The effective metric  components for an arbitrary Lovelock theory are obtained using similar methods to those used in the previous examples (recall that $k_{max}=\frac{D-1}{2}$),
\begin{align}
{\big[G^{tt}\big]^{\widetilde{\textbf{a}}}_{\widetilde{\textbf{c}}} } &= g^{tt}\left(1+\sum_{k=2}^{k_{max}}\dfrac{(D-3)!}{(D-2k-1)!}\lambda_k(2H^2)^{k-1}\right),
\label{frTT1}
\\
{\big[G^{\bar{b}\bar{d}}\big]^{\widetilde{\textbf{a}}}_{\widetilde{\textbf{c}}} } &= g^{\bar{b}\bar{d}} \left(1+\sum_{k=2}^{k_{max}}\dfrac{(D-3)!}{(D-2k-1)!}\lambda_k(2H^2)^{k-1}\left[ 1+\dfrac{2(k-1)}{D-3}\frac{\dot{H}}{H^2}\right]\right)~.
	\label{frSS1}
\end{align}

We can now see what are the conditions that determine whether $G^{\mu\nu}$ is Lorentzian. If the factor multiplying $g^{tt}$ in Eq.~(\ref{frTT1}) is positive, for example if $\lambda_k>0$ for all $k$, then the conditions are similar to the ones found in the previous discussion. The metric can become non-Lorentzian when the $\lambda_k H^2$ is large for at least  some $\lambda_k$ and the factor $1+\dfrac{2(k-1)}{D-3}\dfrac{\dot{H}}{H^2}$ is negative. A simple example is provided by the case when $\lambda_{k_{max}}>0$ is the dominant coupling and all the rest are small. Then if $H^2+\dot{H}<0$, the effective metric is non-Lorentzian. This condition means that for solutions of the form $a(t)\sim t^\alpha$, all the range of decelerated expansion $0 < \alpha < 1$ leads to non-Lorentzian metric.

Since for violations of hyperbolicity one needs both $\dot H$ and $H$ to be large, it follows that $H$ needs to change rapidly in time; therefore at some late time, one expects that $H$ becomes small and then the correction terms are no longer significant. Alternatively $H$ becomes so large that the semiclassical approximation breaks down.

Our interpretation of the results is that Lovelock theories cannot be viewed as  consistent theories of modified gravity.

\subsection{Spherically symmetric  black holes}\label{s3.2}

In this subsection we only discuss simple examples to illustrate the applicability of the method also for static solutions and in order to expose the similarities and differences with respect to the  discussion of cosmological backgrounds. This will enable us to compare our results to the results obtained by other methods. So, we perform the calculation for the simplest cases, GB in 5D and 6D for static BH solutions. Extending the calculations to more complicated cases is straightforward.

The static spherically symmetric BH solutions for Lovelock theories take the standard form \cite{Boulware:1985wk, Cai:2003gr, Wheeler:1985nh}.
\begin{equation}
ds^2=-f(r)dt^2+f(r)^{-1}dr^2+r^2d\Omega_{D-2}^2,
	\label{BHmet}
\end{equation}
where $d\Omega_{D-2}^2$ is the standard metric of the $D-2$ unit sphere. The metric is a solution of the EOM when $F(r)=-f(r)/r^2 $ is the solution of the polynomial equation   \cite{Myers:1988ze},
\begin{gather}
\sum_{n=2}^{k_{max}}\bigg(\lambda_n\bigg[\prod_{l=1}^{2n-2}(D-l-2)\bigg]F(r)
^n\bigg)+F(r)=\dfrac{M}{r^{D-1}}~.
			\label{BHf(r)}
\end{gather}
Here we set $\Lambda=0$, $16\pi G=1$ and the parameter $M$ is referred to as the black hole mass.

The non-vanishing components of the Riemann components are the following,
\begin{eqnarray}
		& \mathcal{R}^{tr}_{tr} =-\dfrac{f''(r)}{2}~,
		\label{f1}
	\\ &
		\mathcal{R}^{ij}_{kl}=-\dfrac{f(r)}{r^2}\delta^{ij}_{kl}~,
		\label{f2}
\\ &
		\mathcal{R}^{~\alpha i}_{\alpha j}=-\dfrac{f'(r)}{2r}\delta^i_j.
		\label{f3}
\end{eqnarray}
The indices  $i, j, k, l=1,2,\dots D-2$ denote angular coordinates and $\alpha = t, r$. The main difference compared to the FRW case is that there are three different kinds of non-vanishing Riemann tensor components as compared to two non-vanishing components in the FRW case. Since the Riemann tensor components only depend on $f$, $f'$ and $f''$ but not on higher derivatives of $f$, the effective metric will be  a multinomial in these quantities.

Gauge-invariant tensor perturbations are defined in the standard way by
\begin{subequations}
\begin{align}
	h^{\alpha\beta} & =0,
	\label{g11}
\\
h^{\alpha i} & =0,
	\label{g22}
\\
h^{i}_{~i} & =0,
	\label{g13}
\\\nabla_i h^{ij} & =0~.
	\label{g24}
	\end{align}
\end{subequations}
Recall that such tensor perturbations can always be defined when expanding about a maximally symmetric space.

For the 5D GB, the solution can be found using Eq.~(\ref{BHf(r)}),
\begin{equation}
f_5(r)=1+\dfrac{r^2}{4\lambda_2}\bigg(1-\sqrt{1+\dfrac{16\lambda_2 M}{r^4}}\bigg)~.
			\label{BH5Dsol}
\end{equation}
As mentioned previously, negative $\lambda_2$ is problematic for the following reason. The GB term is significant when $\dfrac{16\lambda_2 M}{r^4} \gtrsim 1$. But, when $\lambda_2<0$,  the  metric has a branch cut at $16|\lambda_2| M/r^4=1$. So the metric makes sense only for cases when the GB term is subdominant. This makes this case uninteresting, since we know that when the GB term is subdominant to the Einstein term, the effective metric for perturbation is Lorentzian. So, we will focus on the case that $\lambda_2>0$.

The calculation of the effective metric is carried out in a similar way to the one performed in the previous FRW examples.  The effective metric components are given by
\begin{gather}
\left[G^{\alpha\alpha}\right]^{\widetilde{\textbf{i}}}_{\widetilde{\textbf{j}}}= g^{\alpha\alpha}\left(1-2\lambda_2\dfrac{f_5'(r)}{r}\right)~,		
	\label{BH5}\\
\left[G^{kk}\right]^{\widetilde{\textbf{i}}}_{\widetilde{\textbf{j}}} =g^{kk}\Bigl(1-2\lambda_2 f_5''(r)\Bigr)~.
			\label{BHef5}
\end{gather}
As in the FRW case, symmetry results in an equal effective metric for all polarizations.

For the effective metric $G^{\mu\nu}$ not to be Lorentzian several conditions have to be satisfied. First, it is clear that for any value of $\lambda_2$, for large enough values of $r$, the metric is approximately that of a Schwarzschild solution of Einstein's equations, so the effective metric will be Lorentzian in this region of large $r$. Therefore, effective metric can be non-Lorentzian only for smaller values of $r$. Then, the part of the effective metric corresponding to the $\alpha\alpha$ components is Lorentzian since it is proportional to the metric $g^{\alpha\alpha}$.

Since the metric in Eq.~(\ref{BH5}) is always Lorentzian, there exist two possibilities:\\
A) The global sign of the metric $G^{\mu\nu}$ does not affect the
hyperbolicity of the EOM. In this case,  the global sign of Eq.~(\ref{BHef5}) does not affect the hyperbolicity of the EOM (this is also the case in 6D example discussed below.)\\
B) The global sign of $G^{\mu\nu}$ affects the hyperbolic character of
the EOMs. Then, since Eq.~(\ref{BH5}) is Lorentzian, the hyperbolicity is
determined by the sign of the factor in parentheses in Eq.~(\ref{BHef5}).
We conclude that for the effective metric not to be Lorentzian the corrections have to be large
and such that $2 \lambda_2 f_5^{\prime\prime}(r)> 1$.


The following picture emerges: There could be some range in $r$ where the corrections due to the GB term are large. In this range, the hyperbolicity is governed by the higher order terms and it is not guaranteed. Additional conditions must be imposed such that the effective metric is indeed Lorentzian. Therefore, generically, we do not expect a Lorentzian effective metric in this limited range. However, for the 5D case, it turns out that for the solution Eq.~(\ref{BH5Dsol}),  the corrections due to the GB term are never large enough and so the effective metric is Lorentzian everywhere outside the horizon.

We proceed by considering the 6D solution,
\begin{equation}
f_6(r)=1+\dfrac{r^2}{6\lambda_2}\bigg(1-\sqrt{1+\dfrac{20\lambda_2M}{r^5}}\bigg)
			\label{BHD6}~.
\end{equation}
Similar to the 5D example, the case $\lambda_2<0$ is uninteresting, so we focus on the case $\lambda_2>0$.

The effective metric components in this case are
\begin{gather}
\left[G^{\alpha\alpha}\right]^{\widetilde{\textbf{i}}}_{\widetilde{\textbf{j}}}= g^{\alpha\alpha}\Bigg(1-4\lambda_2\dfrac{rf_6 '(r)+f_6(r)-1}{r^2}\Bigg)~,\\
\left[G^{kk}\right]^{\widetilde{\textbf{i}}}_{\widetilde{\textbf{j}}} =g^{kk}\Bigg(1-\lambda_2\dfrac{2rf_6''(r)+4f_6'(r)}{r}\Bigg)~,
			\label{BH-H6}
\end{gather}
and the hyperbolicity condition reads
\begin{gather}
\lambda_2\dfrac{2rf_6''(r)+4f_6'(r)}{r}<1~.
			\label{BH-H6con}
\end{gather}
The six dimensional $f_6(r)$ satisfies $f_6''(r)<0$, $f_6'(r)>0$, $f_6(r)>0$. Thus, the effective metric is non-Lorentzian for  certain values of $\lambda_2$ depending on the $r$ and $M$. Here the range in which the metric is non-Lorentzian is $r\lesssim (2.26 M\lambda_2)^{1/5}$. In order to have  $r \approx (2.26 M\lambda_2)^{1/5}$ outside the horizon $\lambda_2$ has to be large enough, $ \lambda_2 \gtrsim (5.6 M)^{2/3}$.  In this case there is a region which extends from the horizon to some maximal radius in which the Einstein solution is substantially modified and in which the effective metric is not Lorentzian. One can also view this constraint as setting an effective cutoff length scale such that the corrections are subdominant.

\end{subequations}

\section{Summary and Conclusion}

In this paper we addressed the question whether Lovelock gravity can constitute a truly modified theory of gravity. We found that it cannot. Our conclusion is based on analyzing the hyperbolicity of the EOM of perturbations around solutions of Lovelock gravity. We found that, generically, if the Lovelock correction terms are comparable to or larger than the Einstein term, then the EOM of perturbations are not hyperbolic and therefore the EOM are not causal in this case. When the Lovelock terms are small and therefore provide only a perturbative correction to the Einstein term, then the EOM are hyperbolic because they are hyperbolic for GR.

We calculated the effective metric in field space for Lovelock theories by generalizing the method for such calculations in scalar field models.  The new formalism developed in Sec.~(\ref{s2.2}) enabled us to identify the effective metric for Lovelock theories and determine the conditions for  hyperbolicity of the perturbed EOM about different backgrounds.

Then, we performed explicit calculations of the effective metric in some examples. First, we considered FRW cosmological solutions in Sec.~(\ref{s3.1}). We found that the hyperbolicity of the effective metric is governed by the magnitude of $\lambda_2 H^2$, and the sign of $\dot{H}$. The result is that when $\lambda_2 H^2$ is large and $\dot{H}$ is negative, the effective metric is not Lorentzian. Specifically, when the highest Lovelock term is dominant, the whole range of decelerated expansion $a(t)\sim t^{\alpha}$, $0<\alpha<1$ leads to non-Lorentzian effective metric. Our results reproduce the results of Papallo and Reall \cite{Papallo:2017qvl} that were obtained by the more general method of characteristics. It follows that the cutoff scale of the theory is not set by the Planck scale or some other independent high scale; rather it set by the correction terms, ensuring that the correction terms are subdominant. If the correction terms are subdominant, the EOM of perturbations are hyperbolic because the EOM of perturbations in GR are hyperbolic.

Our discussion ended with an investigation of some simple spherically symmetric BH solutions. This was performed to show that one can apply our formalism also to this case.
We considered only the Einstein-Gauss-Bonnet theory in 5D and 6D.
The effective metric is found to be non-Lorentzian only in 6D over the range $r\lesssim(2.26M\lambda_2)^{1/5}$ which is located outside the horizon when $\lambda_2 \gtrsim(5.6M)^{2/3}$, in agreement with \cite{Papallo:2015rna}.
Our results indicate that Lovelock theories lead to EOM for perturbation  which are not hyperbolic and thus imply causality violations in agreement with the results of \cite{Camanho:2014apa} and  \cite{Benakli:2015qlh}. Again, the conclusion is that the cutoff scale of the theory is  set by the correction terms, ensuring that the correction terms are subdominant and the EOM of perturbations are hyperbolic.

Looking ahead, the effective metric approach for studying causality violations can also be implemented for modified gravity theories other than Lovelock,  such as $F(\mathcal{R})$, $k$-essence etc.

Another interesting direction is to
investigate the relation between the hyperbolicity of
the EOM to the conditions under which one can have a perturbative treatment of Lovelock theory \cite{MenaMarugan:1990ji,Simon:1990ic}
and define a Hamiltonian for higher-derivative gravity theories \cite{Teitelboim:1987zz,MenaMarugan:1992if}.
These are found to be closely connected for scalar field models as
shown in \cite{Aharonov:1969vu} and so are expected to be related also for Lovelock theories.

\section*{Acknowledgments}
We would like to thank Gary Gibbons for discussions, Alex Vikman for many valuable comments on the manuscript, Giuseppe Papallo for help with comparison to [22] and especially Harvey Reall for useful discussions, comments and suggestions.  The research was supported by the Israel Science Foundation grant no. 1294/16.

\end{document}